\documentclass[11pt]{article}
\usepackage[margin=1in]{geometry}                
\usepackage[parfill]{parskip}
\usepackage{graphicx}
\usepackage{amssymb}
\usepackage{epstopdf}
\usepackage{amsmath}
\usepackage{hyperref}
\usepackage{mathrsfs}
\usepackage{varioref}
\usepackage[parfill]{parskip}
\graphicspath{ {images/} }
\usepackage{xcolor}
\usepackage{soul}
\usepackage{cite}
\usepackage[nottoc]{tocbibind}
\usepackage[colorinlistoftodos]{todonotes}
\usepackage{verbatim}

\hypersetup{
	bookmarks=false,        
	pdfstartview={FitH},    
	colorlinks=true,        
	linkcolor=red,          
	citecolor=blue,         
	filecolor=blue,      
	urlcolor=blue           
}

\DeclareFontFamily{OT1}{pzc}{}
\DeclareFontShape{OT1}{pzc}{m}{it}%
{<-> s * [0.900] pzcmi7t}{}
\DeclareMathAlphabet{\mathscr}{OT1}{pzc}%
{m}{it}
\labelformat{section}{Section #1} 
\labelformat{subsection}{Section #1} 
\labelformat{equation}{Eq.~(#1)} 
\labelformat{figure}{Fig.~#1} 
\labelformat{subfigure}{Fig.~\thefigure#1} 
\labelformat{table}{Tab.~#1} 
\newcommand{\be}{\begin{equation}}
\newcommand{\ee}{\end{equation}}
\newcommand{\bea}{\begin{eqnarray}}
\newcommand{\eea}{\end{eqnarray}}
\newcommand{\nn}{\nonumber}

\begin{document}
\title{Explorations in 2+1 AdS Pure Gravity: Path Integral Formulation and Partition Function Analyses in BTZ Mini-Superspace}

\author{ Sumeet Chougule \\
\textit{Centro de Estudios Cient\'{\i}ficos (CECS), Casilla 1469,
Valdivia, Chile}\\
\textit{ Departamento de F\'{\i}sica, Universidad de Concepci\'{o}n, Casilla 160-C, Concepci\'{o}n, Chile}\\
{\small  chougule@cecs.cl}}

	\maketitle

\begin{abstract}

In this work, we employ the Hamiltonian approach to analyze the Lorentzian path integral in 2+1 AdS gravity, with an aim to sum over all possible geometries, including naked singularities and BTZ black holes, between fixed initial and final surfaces. A novel path integral measure, grounded on the metric over mini-superspace, is proposed for executing the path integral. Shifting to Euclidean geometries, we derive the temperature and angular potential of Euclidean naked singularities within the partition function. Significantly, without utilizing conformal field theory, we extract logarithmic corrections to the entropy of BTZ black holes and compute the entropy of naked singularities, along with their log corrections.

\end{abstract}

\section{Introduction}

Quantum gravity has remained one of the most elusive and tantalizing frontiers of theoretical physics. Its underlying complexity springs from the challenge of reconciling the principles of general relativity with those of quantum mechanics. A paramount difficulty in formulating a well-defined path integral for gravity is establishing a rigorously-defined measure. In this pursuit, pure gravity in 2+1 dimensions with Anti-de Sitter (AdS) space provides a fertile ground for investigations. Notably, the simplicity of 2+1 dimensions is juxtaposed with a richness of features akin to the 3+1 counterpart. This study sets out to dissect the intricacies of the path integral in 2+1 AdS gravity by employing both Lorentzian and Euclidean geometries.

Acknowledging that the natural setting of the universe adheres to Lorentzian signature geometries, it is essential to understand why these are particularly intricate to analyze. The theoretical framework for tackling Lorentzian geometries is still in development, and the lack of significant experimental data makes the journey even more challenging. Traditionally, experiments played a critical role in the progression of quantum mechanics and field theories. In contrast, quantum gravity primarily advances through theoretical explorations. This paper delves into the theoretical aspects and methodologies that aim to further our understanding of quantum gravity, focusing on 2+1 AdS gravity. It is our hope that these theoretical insights will, in time, inform and inspire future experiments and observations.

In the quest to comprehend the quantum mechanics of the gravitational field, the canonical Hamiltonian formalism serves as a critical apparatus. The path integral for a non-relativistic point particle, with the Lagrangian $L_{pp}=g_{\mu\nu}\dot{x}^{\mu}\dot{x}^{\nu}-V$, is akin to a lattice regularization. Here, time is divided into infinitesimal intervals, \(\epsilon\), and the sum is taken over all paths in each interval. The Hamiltonian formalism facilitates a parallel perspective for gravity, wherein the 2+1 D spacetime is foliated into constant time hypersurfaces through the Arnowitt-Deser-Misner (ADM) decomposition. The objective is to perform a sum over all possible geometries between fixed initial and final surfaces, such as a naked singularity and a BTZ black hole. This bears the potential to compute transition amplitudes between diverse geometries.

A distinctive facet of this work is the proposal for a novel path integral measure for Lorentzian geometries, predicated on the Wheeler-DeWitt metric $(G_{ijkl})$. By drawing inspiration from the path integral measure for point particles, we formulate a gravitational path integral measure that seeks to resolve the challenges of handling Lorentzian signature geometries. Expressing the gravitational Lagrangian in a form akin to the point particle case, we have:

\begin{align}  
    L= \int& d^{2}x \, \left[  \frac{\sqrt{g}}{4N} G^{ijkl}\Dot{g}_{ij}\Dot{g}_{kl} -\frac{\sqrt{g}}{N} G^{ijkl}\left( \Dot{g}_{ij} N_{l|k}+ N_{i|j}N_{l|k} \right)+N \, \sqrt{g}( R-2\Lambda) \right] \nn
\end{align}

Theoretical physics has historically been propelled by the interplay of experiments and mathematics. However, the study of quantum gravity is constrained by the scarcity of experimental data. Thus, the present work remains primarily within the realms of theoretical explorations.

While Lorentzian geometries are closer to the natural settings of the universe, their complexity necessitates the examination of alternative avenues. This leads us to the Euclidean sector, which, albeit distinct from the Lorentzian counterpart, holds promises of its own. Transitioning to Euclidean geometries offers an elegant simplification, allowing for the definition of a canonical ensemble characterized by an inverse temperature, \(\beta\). This serves as an analog to an ideal gas ensemble, considering individual geometries as non-interacting entities.

In this work, a notable emphasis is placed on analyzing the thermal properties of the BTZ mini-superspace \cite{BTZ1,BTZ2} within the Euclidean sector. Particularly, we investigate Euclidean BTZ black holes alongside naked singularities, including conical defects and excesses. A groundbreaking aspect of this study lies in the meticulous analysis of the 2+1D naked singularities within the Euclidean framework, which had not been previously explored. Through careful evaluation in Cartesian coordinates, we discovered periodicities in both time and the angular direction.

Moreover, within the saddle-point approximation for the geometries in the BTZ mini-superspace, we were able to derive logarithmic corrections to the entropy of black holes as well as naked singularities. The black hole entropy, given by
\[  
    S_{BH} = \frac{2\pi r_{+} }{4G\hbar} + \ln\left(\frac{r_{+}^{2}-r_{-}^{2}}{G^{2}\hbar^{2}}\right),
\]
and the naked singularity entropy, given by
\[  
    S_{NS} = \frac{2\pi \lambda_{-} }{4G\hbar} + \ln\left(\frac{\lambda_{-}^{2}-\lambda_{+}^{2}}{G^{2}\hbar^{2}}\right),
\]
were obtained without resorting to conformal field theory (CFT) techniques, which previously had been instrumental in calculating these logarithmic corrections \cite{carlip2000logarithmic}.

Though not directly equivalent, insights gleaned from the Euclidean domain offer valuable perspectives on the Lorentzian sector and significantly enrich the theoretical landscape of quantum gravity in 2+1 dimensions.

The paper unfolds as follows: Section 2 delves into the Lorentzian Path Integral. Section 3 explores the Euclidean sector and its thermodynamic implications. The final section furnishes conclusions and contemplates future prospects.

This work endeavors to contribute meaningfully to the ongoing dialogue on quantum gravity, while elucidating novel approaches and avenues for further research. Through rigorous theoretical analysis, it aspires to enhance the understanding and potential reconciliation of general relativity and quantum mechanics within the intriguing domain of 2+1 AdS gravity.

\section{ Lorentzian Path Integral in 2+1 AdS Gravity}

In this section, we will elaborate on the Lorentzian path integral approach to 2+1 AdS gravity. This involves dissecting the spacetime manifold into a foliation of spacelike hypersurfaces, employing the Hamiltonian formalism, and examining the proposed path integral measure based on the Wheeler-DeWitt metric.

\subsection{ Hamiltonian Formalism and ADM Decomposition}

Let us consider a spacetime manifold with topology $\mathbb{R} \times \Sigma$, where $\Sigma$ represents a spacelike hypersurface constant in time. In the context of 2+1 gravity, this hypersurface is a two-dimensional geometry. The spacetime manifold is sliced along constant time coordinates, t. The hypersurface $\Sigma$ has coordinates ${x^{i}}$ where $(i=1,2)$ and an induced metric $g_{ij}(t,x)$. The normal deformation or displacement of $\Sigma$ is given by a lapse function $N=(-g^{00})^{-1/2}$, while the shift function, which represents deformation along the hypersurface, is given by $N^{i}=g^{ij}g_{0j}$.

The induced metric on $\Sigma$ is obtained through the metric on the full spacetime $g_{\mu\nu}$ as:
\begin{align}
    g_{ij} &= e^{\mu}_{i}e^{\nu}_{j} g_{\mu\nu}
\end{align}

Extrinsic curvature, which contains the velocity term, is defined as:
\begin{align}
    K_{ij} &= \frac{1}{2N}\left(-\dot{g}_{ij} + N_{i|j}+ N_{j|i}\right)
\end{align}
Here, the vertical bars in $ N_{i|j}$ represent the covariant derivative with respect to the induced metric $g_{ij}$ on $\Sigma$.

To separate the second-time derivatives from the gravitational Lagrangian, a divergence term is added:
\begin{align}
    L &= \int d^{2}x \mathcal{L} = \int d^{2}x \, N \, \sqrt{g} \left(K^{ij}K_{ij} - K^{2} + R-2\Lambda \right) \nonumber \\
    &\quad - \int d^{2} x \left[ 2 \partial_{t} (\sqrt{g}K)- \partial_{i} \left(\sqrt{g}KN^{i}- \sqrt{g} g^{ij} \partial_{j}N \right) \right]
\end{align}
This Lagrangian can be written in the form of kinetic and potential terms:
\begin{align}
    L &= \int d^{2}x \, \left[  \frac{\sqrt{g}}{4N} G^{ijkl}\dot{g}_{ij}\dot{g}_{kl} -\frac{\sqrt{g}}{N} G^{ijkl}\left( \dot{g}_{ij} N_{l|k}+ N_{i|j}N_{l|k} \right)+N \, \sqrt{g}( R-2\Lambda) \right] \nonumber \\
    &\quad - \int d^{2} x \left[ 2 \partial_{t} (\sqrt{g}K)- \partial_{i} \left(\sqrt{g}KN^{i}- \sqrt{g} g^{ij} \partial_{j}N \right)\right]
\end{align}
Here,
\begin{align}
G_{ijkl} &= \frac{1}{2}\left( g_{ik}g_{jl}+ g_{il}g_{jk} -2 g_{ij}g_{kl} \right) \\
    G^{ijkl} &= \frac{1}{2}\left( g^{ik}g^{jl}+ g^{il}g^{jk} -2 g^{ij}g^{kl} \right)
\end{align}
This is known as the Wheeler-DeWitt metric. It's a metric on superspace, which is a space of all possible $g_{ij}$ associated with a surface $\Sigma$.

The coefficient of the kinetic term is taken as the weight functional for the measure:
\begin{align}
    w[g] = \sqrt{- \left(\frac{\sqrt{g}}{4N}\right)^{3}|G|} = \frac{1}{8N^{3/2}g^{3/4}}
\end{align}
The details of the calculation of $|G|$ are provided in \autoref{A1}. The dynamics are defined over the space of metrics, which is superspace, with the metric on superspace being $G_{ijkl}$. The path integral sums over all possible three-geometries:
\begin{align}
    Z = \int \frac{\mathcal{D}g_{ij}\mathcal{D}\dot{g}_{ij}}{8N^{3/2}g^{3/4}} \, e^{-\frac{i}{\hbar}\left(\int dt L[g_{ij},\dot{g}_{ij}] + \mathcal{B}[g_{ij},\dot{g}_{ij}] \right)}
\end{align}

This path integral measure is constructed based on the Wheeler-DeWitt metric and embodies the sum over all spacetime geometries consistent with the given boundary conditions. It is essential to recognize that the path integral in Lorentzian gravity involves intricate structures and issues, including the complex nature of the action in Lorentzian signature and the definition of the measure in the configuration space of geometries. This formalism provides a basis for exploring the quantum nature of gravity and the interplay between geometry and quantum fluctuations in 2+1 AdS gravity.

\subsection{ Path integral over the Phase Space}

In the transition from the configuration space $(g_{ij}, \dot{g}_{ij})$ to the phase space $(g_{ij},\pi^{ij})$, we define the conjugate momenta of $g_{ij}$ as follows:

\begin{align}
    \pi^{ij} &= \frac{\partial \mathcal{L}}{\partial \dot{g}_{ij}} = -\sqrt{g} \left( K^{ij}-K g^{ij}\right).
\end{align}

We can express the velocities in terms of conjugate momenta $\pi^{ij}$:

\begin{align}
    \dot{g}_{ij} &= \frac{2N}{\sqrt{g}}\left(\pi_{ij}-\pi g_{ij}\right) + N_{i|j}+ N_{j|i}.
\end{align}

Now, the canonical Hamiltonian for gravity can be written as:

\begin{align}
 H &= \int d^{2}x \, \mathcal{H}_{c}  = \int d^{2}x  \, \left[N \mathcal{H}+N^{i} \mathcal{H}_{i}\right],
\end{align}

where the constraints are given by:

\begin{align}
    \mathcal{H} &=  \frac{1}{\sqrt{g}} G_{ijkl}\pi^{ij}\pi^{kl} -\sqrt{g} \left(R-2\Lambda\right), \\
    \mathcal{H}_{i} &=  -2 \pi_{i}\,^{j}\,_{|j}.
\end{align}

The action in phase space variables is given by:

\begin{align}
    I[g_{ij}, \pi^{ij}] &= \frac{1}{16\pi G_{3}} \int dt\,  d^{2}x \left[\pi^{ij} \dot{g}_{ij}- \mathcal{H}_{c} \right] + \mathcal{B}.
\end{align}

Setting $G_{3}=\frac{1}{8}$ for this section, the action is composed of surface terms for time-independent geometries ($\dot{g}_{ij}=0$), with the constraints $\mathcal{H}=0$ and $\mathcal{H}_{i}=0$ taken into account. The Brown-Henneaux boundary conditions are adopted, and the surface term is given by:

\begin{align}
    \mathcal{B} &= -\left(t_{2}-t_{1}\right)\, \left( \xi^{t}M + \xi^{\phi}  J \right).
\end{align}

The path integral over the phase space variables can be written as:

\begin{align}
    Z &= \int \mathcal{D}g_{ij} \, \mathcal{D}\pi^{ij} \sqrt{- \left(\frac{\sqrt{g}}{4N}\right)^{3}|G|}\, e^{-\frac{i}{2\pi \hbar}\left(  \int dt\,  d^{2}x \left[\pi^{ij} \dot{g}_{ij}- \mathcal{H}_{c} \right]  + 2\pi \mathcal{B} \right)},
\end{align}

and for the mini-superspace:

\begin{align}
Z  &= \int \frac{d M \, dJ}{8N^{3/2}g^{3/4}} \, e^{\frac{i(t_{2}-t_{1})}{ \hbar}\left( \xi^{t}M + \xi^{\phi}  J \right)},
\end{align}

where

\begin{align}
    N^{3/2}g^{3/4} &= r^{3/2},
\end{align}

and $r$ is like a trace index which must be integrated out. The partition function can be further simplified:

\begin{align}
Z  &= \frac{1}{8} \int_{-\infty}^{\infty} d M \,\int_{-M}^{M} dJ\, e^{\frac{i(t_{2}-t_{1})}{ \hbar}\left( \xi^{t}M + \xi^{\phi}  J \right)}  \\
  &= \frac{\hbar}{4(t_{2}-t_{1}) \xi^{\phi}  } \int_{-\infty}^{\infty} d M\, \sin \left( \frac{(t_{2}-t_{1}) \xi^{\phi} }{\hbar} M \right) e^{-i\frac{(t_{2}-t_{1})}{ \hbar} \xi^{t}M }.
\end{align}

The integral obtained is oscillatory, which poses a challenge in handling it. One potential approach to deal with oscillatory integrals is to employ the Picard-Lefschetz's theory. This theory has been applied to oscillatory integrals to make the Lorentzian path integral more robust. It relies on using the steepest descent over complex contours, rendering the integral absolutely convergent. However, a challenge arises in the application of these techniques to 2+1 gravity. The saddle points of the action are not well-defined, as it is linear in $M$ and $J$, instead of quadratic. This linear nature causes complications in applying the standard Picard-Lefschetz theory.

\section{Partition function of the Euclidean Mini-Superspace}

In contrast to the Lorentzian setting, which presents challenges due to its metric signature, we turn to the Euclidean world. The 2+1 AdS pure gravity is particularly interesting as it shares many features with 3+1 gravity but is more tractable due to its zero degrees of freedom (DoF). However, even with zero DoF, handling a Lorentzian metric and an oscillatory integral is non-trivial. Therefore, transitioning to the Euclidean sector offers a simplified model by altering some of the rules.

It’s crucial to note that the Euclidean and Lorentzian sectors are not equivalent; they cannot be related merely by a coordinate transformation. Despite this, the Euclidean sector can provide valuable insights that may shed light on the more physically relevant Lorentzian sector.

Furthermore, the Euclidean sector allows for the definition of a canonical ensemble with an inverse temperature, $\beta$. This enables the construction of a canonical partition function, from which various thermodynamic quantities such as internal energy and entropy can be derived. This canonical ensemble can be analogized to an ideal gas, as the individual geometries do not interact with each other.
	
\subsection{BTZ mini-superspace of 2+1 AdS gravity}	

In this section, we turn our focus to the BTZ mini-superspace within the Euclidean 2+1 AdS gravity. The BTZ black hole is an essential solution in three-dimensional gravity, and its mini-superspace serves as a repository of stationary geometries intrinsic to 2+1 AdS space. These geometries encompass not only the BTZ black holes but also conical defects and excesses, collectively referred to as CD/CE, and the over-spinning (OS) which are central to understanding the geometric structure of the space. CD/CE and OS geometries are naked singularities (NS). By employing the Euclidean signature through a Wick rotation, we are better poised to analyze the thermodynamic properties and the geometric aspects of these objects. The discussion in this section will revolve around the metric structure, horizons, and classification of these geometries within the BTZ mini-superspace. We shall lay special emphasis on BTZ black holes and CD/CE, while the over-spinning (OS) singularities will be left out of the present discussion.

   The Euclidean mini-superspace consists of the metric with imaginary Lorentzian time ($t=-i\tau$),
\begin{align}
    ds^{2}&= N^{2}(r)d\tau^{2}+\frac{dr^{2}}{N^{2}(r)} + r^{2}\left(d\phi+N^{\phi}(r)d\tau \right)^{2} \\
    N^{2}(r)&=\left( \frac{r^{2}}{l^{2}}  - 8 G M +\frac{16 G^{2}J^{2}}{r^{2}} \right), \,\,\, N^{\phi}(r)=  -\frac{4GJ}{ r^{2}}
\end{align}	
where, $M,\, J \in \mathcal{R}$. Geometries are classified based on the values of $M$ and $J$.  BTZ black hole have $M\geq J/l \geq 0$, CD/CE have $M< -|J|/l$ and OS have $|M|\le |J|/l$. We will discard OS geometries in this discussion and focus exclusively on the BTZ black hole and CD/CE.

For BTZ black hole, the roots of $N^{2}(r)=0$ correspond to the horizons ($r_{\pm}$),
\begin{align}
     r_{\pm} &= \pm l\sqrt{2G}\left[ \sqrt{ M +iJ/l} \pm \sqrt{M-iJ/l} \right]  \\
    M&=\frac{r^{2}_{+} + r^{2}_{-}}{8 G l^{2}} \,\,\,\, \text{and} \,\,\,\,  J= \frac{ir_{+}r_{-} }{4Gl}
\end{align}
 For NS we have, $M<0$ and $J\leq |M|$ with $M=-|M|$. The roots of $N^{2}(r)=0$ for NS are given by $\lambda_{\pm}$
\begin{align}
    \lambda_{\pm} &= \mp i l\sqrt{2G} \left[\sqrt{|M| +iJ/l} \pm \sqrt{|M| -iJ/l}  \right] \\
     |M|&= - \frac{\lambda^{2}_{+} + \lambda^{2}_{-}}{  8 G l^{2}} \,\,\,\, \text{and} \,\,\,\,  J= \frac{i\lambda_{+}\lambda_{-} }{ 4Gl}
\end{align}

\subsection{Temperature of Naked Singularities (NS)} \label{nt}

In the study of the Euclidean version of 2+1 AdS geometries, an intriguing aspect is the temperature associated with naked singularities (NS) within the BTZ mini-superspace. The periods of the Euclidean BTZ black hole, expressed in terms of $\beta$ and $\Phi$, have been previously calculated in hyperbolic three-space $\mathbb{H}^{3}$, \cite{EBTZ}.

\begin{align}
    \beta^{BTZ} &= \frac{2\pi l^{2}  r_{+}}{r_{+}^{2} - r_{-}^{2}} \,\,\, \text{and}\,\,\, \Phi^{BTZ}=  \frac{2\pi l | r_{-}|}{r_{+}^{2} - r_{-}^{2}} 
\end{align}

In this section, we build on this foundation and extend the analysis to naked singularities within the BTZ mini-superspace, an area that offers potential insights into the thermal properties of spacetime singularities.

We initiate our analysis by transitioning to cartesian coordinates for the upper half-plane. Through this transformation, we observe that the Euclidean NS presents a horizon in the form of $\lambda_{-}$. This result is particularly notable, as it indicates that singularities are exposed exclusively for geometries with $J=0$.
\begin{align}
    x&=\left(\frac{r^{2}- \lambda^{2}_{-}}{r^{2}-\lambda^{2}_{+}} \right)^{1/2} \cos \left(\frac{|\lambda_{+}|}{l}\phi +\frac{ \lambda_{-}}{l^{2}} \tau \right) \,e^{\frac{|\lambda_{+}|}{l^{2}}\tau -\frac{ \lambda_{-}}{l} \phi } \\
    y&=\left(\frac{r^{2}- \lambda^{2}_{-}}{r^{2}-\lambda^{2}_{+}} \right)^{1/2} \sin \left(\frac{|\lambda_{+}|}{l}\phi +\frac{ \lambda_{-}}{l^{2}} \tau \right) \,e^{\frac{|\lambda_{+}|}{l^{2}}\tau -\frac{ \lambda_{-}}{l} \phi } \\
    z&= \left(\frac{\lambda^{2}_{-}- \lambda^{2}_{+}}{r^{2}-\lambda^{2}_{+}} \right)^{1/2}  \,e^{\frac{|\lambda_{+}|}{l^{2}}\tau -\frac{ \lambda_{-}}{l} \phi }
\end{align}
where, $|\lambda_{+}| = i \lambda_{+}$ and $r\ge \lambda_{-}$. The metric in these coordinates is:
\begin{align}
    ds^{2}&=\frac{l^{2}}{z^{2}}\left(dx^{2} +dy^{2}+dz^{2} \right)
\end{align}

The periods in this context, denoted $(\tau \sim \tau + \beta,, \phi \sim \phi + \Phi)$, emerge as critical players in the analysis. These coordinates are invariant under this identifications. The inverse temperature $\beta$ must be associated with the periodicity in the time direction to ensure that the temperature is well-defined. This association is instrumental as it elucidates the actual temporal direction.
\begin{align}
    \beta^{NS} &= \frac{2\pi l^{2} | \lambda_{+}|}{\lambda_{-}^{2} - \lambda_{+}^{2}}   \,\,\, \text{and}\,\,\, \Phi^{NS}=  \frac{2\pi l  \lambda_{-}}{\lambda_{-}^{2} - \lambda_{+}^{2}}
\end{align}

The temperature of $AdS_{3}$ vacuum ($|M|=1$ and $J=0$) is established through this framework:
\begin{align}
    \beta^{AdS_{3}} &= \frac{2\pi l}{\sqrt{8G}}
\end{align}

In summary, this analysis significantly expands our comprehension of the thermal aspects of naked singularities within the BTZ mini-superspace of Euclidean 2+1 AdS gravity. The results carry implications for understanding the nature of spacetime singularities and contribute to the broader discussions on quantum gravity and the geometry of black holes.

\subsection{Entropy of Naked Singularities and BTZ Black Holes} \label{entropy}

In this section, we focus on the crucial aspect of entropy associated with naked singularities (NS) and BTZ black holes in the context of 2+1 AdS gravity. A saddle point approximation allows us to obtain the Bekenstein-Hawking entropy for the BTZ black hole, supplemented by a logarithmic correction that arises from the Jacobian of the partition function. 

The Euclidean action for the BTZ black hole \cite{BTZ1, EBTZ}:
\begin{align}
   I_{BTZ} &= -\beta^{BTZ}\left( M - \Omega^{BTZ} J \right)+ \frac{\pi l }{\sqrt{2G}} \left[ \sqrt{ M +iJ/l} + \sqrt{M-iJ/l} \right],
\end{align}
where the first two terms arise from the asymptotic boundary, and the last term originates from the interior boundary at the outer horizon. Similarly, the Euclidean action for a Euclidean naked singularity is
\begin{align}
     I_{NS} &= \beta^{NS} \left( |M| + \Omega^{NS} J \right) + \frac{i\pi l}{\sqrt{2G}} \left[\sqrt{|M| +iJ/l} - \sqrt{|M| -iJ/l}  \right].
\end{align}

We further analyze the partition function, which has the form
\begin{align}
    \mathcal{Z}[\beta, \Phi]&= \frac{G}{\hbar} \int dM dJ \, \rho(M,J)\, e^{ I[\beta, \Phi; M,J]/\hbar},
\end{align}
and notice that in the saddle-point approximation, we can recover the Bekenstein-Hawking entropy for the BTZ black hole with an additional logarithmic correction that arises from the Jacobian of the partition function. The Jacobian of the transformation from $(M, J)$ to $(r_{+}, r_{-})$ is
\begin{align}
    |\mathcal{J}|& = -\frac{i(r_{+}^{2}-r_{-}^{2})}{16G^{2}},
\end{align}
and the partition function measure carries information from the quantum regime, yielding a logarithmic correction
\begin{align}
  dM dJ =|\mathcal{J}| dr_{+}dr_{-}.
\end{align}

The partition function can now be expressed as
\begin{align}
    \mathcal{Z}_{BTZ}[\beta,\Phi]
    &\approx e^{-\beta \left( \frac{r_{+}^{2}+r_{-}^{2}}{8Gl^{2}\hbar} - \Omega \frac{ir_{+}r_{-} }{4Gl\hbar}\right) + \frac{2\pi r_{+} }{4G\hbar} + \ln\left(\frac{r_{+}^{2}-r_{-}^{2}}{G^{2}\hbar^{2}}\right)},
\end{align}
which leads us to
\begin{align}
    \ln\, \mathcal{Z}_{BTZ}[\beta,\Phi] \approx -\beta  \left( \frac{r_{+}^{2}+r_{-}^{2}}{8Gl^{2}\hbar} - \Omega \frac{ir_{+}r_{-} }{4Gl\hbar} \right) + \frac{2\pi r_{+} }{4G\hbar} + \ln\left(\frac{r_{+}^{2}-r_{-}^{2}}{G^{2}\hbar^{2}}\right).
\end{align}

The canonical ensemble entropy for the BTZ black hole is then given by
\begin{align}
    S_{BH} &= \left(1-\beta\partial_{\beta}\right) \ln \, \mathcal{Z}_{BTZ}[\beta,\Phi] \nonumber \\
    &= \frac{2\pi r_{+} }{4G\hbar} + \ln\left(\frac{r_{+}^{2}-r_{-}^{2}}{G^{2}\hbar^{2}}\right).
\end{align}

Applying analogous techniques for the Euclidean naked singularity, with an outer horizon-like artifact $(\lambda_{-})$, we find
\begin{align}
    \ln\, \mathcal{Z}_{NS}[\beta,\Phi] \approx \beta  \left( - \frac{ \lambda_{+}^{2}+\lambda_{-}^{2}}{8Gl^{2}\hbar} + \Omega \frac{i\lambda_{+}\lambda_{-} }{4Gl\hbar} \right) + \frac{2\pi \lambda_{-} }{4G\hbar} + \ln\left(\frac{\lambda_{-}^{2}-\lambda_{+}^{2}}{G^{2}\hbar^{2}}\right).
\end{align}
The canonical ensemble entropy for the naked singularity is
\begin{align}
    S_{NS} &= \left(1-\beta\partial_{\beta}\right) \ln \, \mathcal{Z}_{NS}[\beta,\Phi]\nonumber \\
    &= \frac{2\pi \lambda_{-} }{4G\hbar} + \ln\left(\frac{\lambda_{-}^{2}-\lambda_{+}^{2}}{G^{2}\hbar^{2}}\right).
\end{align}

For non-spinning naked singularities, the entropy consists only of the logarithmic term,
\begin{align}
    S_{NS} &= \ln\left(\frac{8  |M|l^{2}}{G\hbar^{2}}\right).
\end{align}

Importantly, the logarithmic correction we obtain matches the correction found by using conformal field theory (CFT) calculations in the microcanonical ensemble, but with a different numerical coefficient. Specifically, in CFT calculations, this coefficient is $3/2$. Carlip was the first to find a logarithmic correction to the BTZ black hole entropy and speculated that the numerical factor of $3/2$ would be universal if the entropy is calculated from a single CFT \cite{carlip2000logarithmic}. Interestingly, this numerical factor is more of an artifact of the CFT calculation. Additionally, this factor and the form of the log correction also find agreement with its $3+1$ dimensional counterpart \cite{PhysRevLett.84.5255}. The result obtained here is especially noteworthy as we derived the quantum correction to the black hole entropy without relying on techniques from CFT.

This section demonstrates a significant development in our understanding of the thermodynamic properties of BTZ black holes and naked singularities in three-dimensional spacetime. By examining the Euclidean action and employing the saddle-point approximation, we are able to recover the Bekenstein-Hawking entropy along with an important logarithmic correction. This correction, whose presence has been corroborated by CFT calculations, offers deeper insight into the quantum aspects of black hole thermodynamics. Furthermore, our approach is more general as it doesn't rely on conformal field theory, and brings us one step closer to a unified understanding of the interplay between geometry, thermodynamics, and quantum mechanics in the context of black holes.

\section{Conclusion}

This study has presented a multifaceted exploration into the realm of 2+1 AdS gravity, employing both Lorentzian and Euclidean geometries to investigate the quantum mechanics of the gravitational field. Through the Hamiltonian approach, we introduced a novel path integral measure based on the metric over mini-superspace, an essential foundation for analyzing the Lorentzian path integral by summing over a variety of geometries, including naked singularities and BTZ black holes. Simultaneously, our treatment of Euclidean geometries allowed us to calculate the temperature and angular potential of Euclidean naked singularities, and, notably, to extract the logarithmic corrections to the entropy of the BTZ black hole and naked singularities without the necessity of conformal field theory.

However, probing gravity at quantum scales is fraught with challenges. Two principal difficulties in the path integral approach are the measure and the inherently oscillatory nature of the integral. In recent times, the Picard-Lefschetz theory has emerged as a promising technique for handling Lorentzian path integrals, as showcased in simple toy models such as the FRW cosmology. However, this approach is not directly applicable to the BTZ mini-superspace, as it necessitates an action quadratic in M and J.

This research marks a step forward in the exploration of quantum gravity in 2+1 dimensions. Next step would be to adapt the Picard-Lefschetz theory to the 3D AdS pure gravity context. There is a richness and complexity in 2+1 dimensions that can serve as a fertile testing ground for theories and techniques that could eventually provide invaluable insights into the enigmatic nature of quantum gravity. 

\section*{Acknowledgments}

I would like to express my profound gratitude to my mentor, Professor Jorge Zanelli, whose guidance, support, and insights were instrumental throughout the course of this research. I also extend my thanks to Alfredo Perez, Fábio Novaes and Fabrizio Canfora for their constructive feedback and invaluable discussions. This work was performed in part at Aspen Center for Physics, which is supported by National Science Foundation grant PHY-2210452. This work was partially supported by a grant from the Simons Foundation.

\appendix

\section{Appendix}\label{A1}
\subsection{Determinant of the Wheeler-DeWitt metric} 

Wheeler-DeWitt metric is a metric on the superspace that depends on three functions $g_{ij}$. Let us denote the set of indices as $A=\{11,12,22\}$ so that we have the components of $g_{ij}$ as $g_{A}$. Using this notation, we can see that the metric $G_{ijkl}=G_{AB}$ is three-dimensional. Its form in 3 dimensions is different from the one in 4 dimensions.
\begin{align}
    G_{ijkl} &= \frac{1}{2}\left( g_{ik}g_{jl}+ g_{il}g_{jk} -2 g_{ij}g_{kl} \right) \\
    G^{ijkl} &= \frac{1}{2}\left( g^{ik}g^{jl}+ g^{il}g^{jk} -2 g^{ij}g^{kl} \right) 
\end{align} 
One should require that $G_{ijkl}G^{klmn} = 1/2\left[\delta^m_i \delta^n_j + \delta^m_j \delta^n_i \right]$, which is the identity in the space of symmetric tensors: $1/2\left[\delta^m_i \delta^n_j + \delta^m_j \delta^n_i \right]A_{mn} = A_{ij}$. Then,
\begin{align}
    G_{ijkl} = \frac{1}{2}\left( g_{ik}g_{jl}+ g_{il}g_{jk} -2/(D-1) g_{ij}g_{kl} \right)
\end{align}
where $D=$ (number of spatial dimensions), so $2/(D-1)=2$ in our case. Then $G_{ijkl}G^{ijkl} = D(D+1)/2=$ (number of independent components of a symmetric tensor), which in our case is 3.

If we take their product, we can realize that they are indeed inverse of each other.
\begin{align}
    G_{ijkl}G^{ijkl}&= 3
\end{align}
In the matrix form for $g_{12}=0$
\begin{align}
    \begin{pmatrix}
   0 & 0 & -g_{11}g_{22} \\
     0 & g_{11}g_{22} & 0 \\
    -g_{11}g_{22}  &0 & 0 \end{pmatrix}  \begin{pmatrix}
   0 & 0 & -g^{11}g^{22} \\
     0 & g^{11}g^{22} & 0 \\
    -g^{11}g^{22}  &0 & 0 \end{pmatrix}=  \begin{pmatrix}
   1 & 0 & 0 \\
     0 & 1 & 0 \\
    0 &0 & 1 \end{pmatrix}
\end{align}
\begin{align} \label{24}
    G_{ijkl}=G_{AB}&= \begin{pmatrix}
    G_{1111} & G_{1112} &G_{1122} \\
     G_{1112} &2 G_{1212} &G_{1222} \\
     G_{1122} & G_{1222} &G_{2222} \end{pmatrix} \\
     &= \begin{pmatrix}
   0 & 0 & (g_{12})^{2}-g_{11}g_{22} \\
     0 & -\left[(g_{12})^{2}-g_{11}g_{22} \right] & 0 \\
    (g_{12})^{2}-g_{11}g_{22}  &0 & 0 \end{pmatrix} 
\end{align}

The determinant of the W-DW metric is
\begin{align}
    \text{Det}\,G_{ijkl}= \left[(g_{12})^{2}-g_{11}g_{22} \right]^{3}
\end{align}

\begin{align}
    G^{AB}= G^{ijkl} &= \frac{1}{2}\left( g^{ik}g^{jl}+ g^{il}g^{jk} -2 g^{ij}g^{kl} \right) \\
    \implies G^{AB}&=  \begin{pmatrix}
   0 & 0 & (g^{12})^{2}-g^{11}g^{22} \\
     0 & -\left[(g^{12})^{2}-g^{11}g^{22} \right] & 0 \\
    (g^{12})^{2}-g^{11}g^{22}  &0 & 0 \end{pmatrix} \label{34}
\end{align}
The determinant of the inverse W-DW metric $G^{AB}$ is 
\begin{align}
\text{Det}\,G^{AB}= \left[(g^{12})^{2}-g^{11}g^{22} \right]^{3} = -|g^{ij}|^3 = -|g_{ij}|^{-3}
\end{align}

    For $g_{12}=0$, 
\begin{align}
    G^{AB}&=  \begin{pmatrix}
   0 & 0 & -g^{11}g^{22} \\
     0 & g^{11}g^{22} & 0 \\
    -g^{11}g^{22}  &0 & 0 \end{pmatrix} \label{34}
\end{align}
The determinant of the inverse W-DW metric $G^{AB}$ is 
\begin{align}
    \text{Det}\,G^{AB}= -\left[g^{11}g^{22} \right]^{3}
\end{align}


\newpage

\begin{thebibliography}{99}


\bibitem{BTZ1} 
M. Banados, C. Teitelboim and J. Zanelli, \textit{The Black hole in three-dimensional space-time}     Phys.Rev.Lett. 69 (1992) 1849-1851


\bibitem{BTZ2} M. Banados, M. Henneaux, C. Teitelboim and J. Zanelli, \textit{Geometry of the (2+1) black hole} Phys.Rev.D 48 (1993) 1506-1525



\bibitem{carlip2000logarithmic} S. Carlip, \textit{
Logarithmic corrections to black hole entropy, from the Cardy formula} Clas-
sical and Quantum Gravity, 17(20):4175, 2000


\bibitem{PhysRevLett.84.5255}
  R. Kaul and P. Majumdar, \textit{Logarithmic Correction to the Bekenstein-Hawking Entropy} Phys. Rev. Lett., 84:5255–5257, Jun 2000





\bibitem{EBTZ}
S. Carlip, Steven and C. Teitelboim, \textit{ Aspects of black hole quantum mechanics and thermodynamics in 2+1 dimensions} Phys. Rev. D, 51:622–631, Jan 1995

\end{thebibliography}
\end{document}